\documentstyle[prl,aps,multicol,epsfig]{revtex}

\newcommand{\equ}{Eq.}

\newcommand{\fig}{Fig.}

\newcommand{\rem}[1]{}

\newtheorem{platztab}{{Tab.}} 
\rem{

}

\newtheorem{platz}{{Fig.}} 
\newcommand{\FIGo}[3]{\begin{figure}%
#3%
\caption[]{\footnotesize #2}%
\label{#1}%
\end{figure}}

\rem{
\newcommand{\FIGo}[3]{
  \begin{center}
    \begin{minipage}{16.0cm}
    \begin{picture}(16.0,1)
      \put(0,0){\framebox(16.0,1){\ }}
    \end{picture}
      \begin{footnotesize}
        \begin{platz}   \label{#1}
          {\rm #2}
        \end{platz}
      \end{footnotesize}
    \end{minipage}
  \end{center}
  }
}

\rem{
\newcommand{\FIGo}[3]{%
\marginpar{\begin{platz} \label{#1} ~ \end{platz} \vspace*{1.5ex} }
}
}

   \def\bege{\begin{equation}}
   \def\ende{\end{equation}}
   \def\bega{\begin{eqnarray}}
   \def\enda{\end{eqnarray}}
   \def\began{\begin{eqnarray*}}
   \def\endan{\end{eqnarray*}}

\newcommand{\weglassen}[1]{}

\newcommand{\todo}[1]{}

\newcommand{\rhogm}{\rho_{\text {gm}}}

\rem{
\addtolength{\topmargin}{-2cm}   
\setlength{\footskip}{1cm}
\setlength{\textheight}{24.5cm}
}
\newcommand{\bstr}{1}

\begin{document}

\title{Singular continuous spectra in a pseudo-integrable billiard}   
\author{Jan Wiersig}
\address{Max-Planck-Institut f\"ur Physik komplexer Systeme,D-01187 Dresden,Germany}
\date{\today}
\maketitle
\begin{abstract}
The pseudo-integrable barrier billiard invented by Hannay and McCraw
[J. Phys. A {\bf 23}, 887 (1990)] -- rectangular billiard with line-segment barrier
placed on a symmetry axis  -- is generalized.  
It is proven that the flow on invariant surfaces of genus two exhibits a
singular continuous spectral component.\\ 
\\
PACS number: 05.45.-a
\end{abstract}

\renewcommand{\baselinestretch}{\bstr} \normalsize

\begin{multicols}{2}
Billiards have attracted much attention in the recent decades as simple
dynamical systems in classical and quantum mechanics. The free motion of a
particle in a domain within a hard elastic boundary shows
a large spectrum of behaviour depending on the shape of the boundary, ranging
from complete regularity to various degrees of chaoticity.

Polygonal billiards are neither chaotic -- the Kolmogorov-Sinai entropy is
zero -- nor integrable in the sense of Liouville-Arnol'd~\cite{Arnold78}
(apart from the rectangles, the equilateral triangles, the $\pi/2, \pi/4,
\pi/4$-triangles and the $\pi/2, \pi/3, \pi/6$-triangles).   
The motion inside a typical polygon is conjectured to be ergodic on the
three-dimensional constant-energy surfaces in phase space, while the motion
inside a rational polygon (all angles are rationally related to $\pi$) is
restricted to two-dimensional invariant surfaces, like in integrable systems,
but the genus of the surfaces is larger than 1. Rational polygonal billiards
are therefore characterized as pseudo-integrable~\cite{RichensBerry81}.   
It is rigorously proven that the flow on these surfaces is ergodic and 
not mixing~\cite{Gutkin96}. Ergodicity means that a particle
with a typical initial momentum explores the entire invariant surface (which
is the billiard table in the configuration space projection). Since
the mixing property is excluded, a small cluster of particles with
identical initial momenta cannot spread out uniformly in time on the invariant
surface.  
But a nonuniform spreading which allows occasional reclustering with decreasing
frequency -- weak mixing in mathematical terms (see e.g.~\cite{AA68}) --
is possible. In fact, weak mixing is believed to be generic, but this is only
shown for polygons all of whose sides are horizontal or vertical.    
Numerical evidence for weak mixing has been reported for the ``square-ring
billiard''~\cite{AGR99}; see also related numerical studies on
rational~\cite{Artuso97} and irrational triangles~\cite{ACG97,CP99}.  

Weak mixing (in the absence of the stronger mixing property) implies
interesting spectral properties~\cite{CFS82}. More precisely, the Fourier
transform of a typical function on the phase space with respect to time has a
fractal-like structure in the frequency domain. Such a spectrum is
characterized as singular continuous, in contrast to discrete spectra of
integrable systems (nonmixing, quasiperiodic motion) and absolutely continuous
spectra of chaotic systems (mixing motion).  
Singular continuous spectra are well known in other fields of physics to
appear at the border between integrability and chaos, e.g. strange nonchaotic
attractors (see, e.g., \cite{PikovskyFeudel94b,FPP96}), the (kicked) Harper
model~\cite{AC94} and models for nonperiodic atomic structures
(see~\cite{GL90} and references therein).  

The spectral properties of a phase space function are related to its
autocorrelation function (AF). For discrete spectra this function is (quasi-)
periodic. In the periodic case the (normalized) AF returns exactly to 1, while
in the quasiperiodic case it comes arbitrarily close to 1. An exponentially
decreasing AF (for increasing time) indicates an absolutely continuous
spectrum. An AF corresponding to a singular continuous spectrum does not
usually decay to zero, but also does not return to 1.  

A nongeneric subclass of pseudo-integrable systems is given by
almost-integrable billiards~\cite{Gutkin86}. A member of this class is,
roughly speaking, composed of several copies of a single completely integrable
billiard, e.g. the $\pi/3,2\pi/3$-rhombus~\cite{EFV84} consists of two
identical equilateral triangles. 
Motion on invariant surfaces of almost-integrable billiards is not weakly
mixing~\cite{Gutkin86}.    
The aim of this letter is to establish that the spectra of these systems
nevertheless can have a singular continuous component (together with a
discrete component). We prove this for a simple model, a generalized version
of the ``barrier billiard''~\cite{HM90}. It is the first time that the
existence of singular spectra in a given billiard is rigorously shown. This is
achieved by decomposing the dynamics into a continuous and a discrete part:
the quasiperiodical motion in the integrable subbilliard and the transitions
between different copies. The latter type of motion, responsible for the
singular continuous component, is discussed with the help of the AF.

We consider a point particle with unit mass elastically bouncing inside a
polygonal enclosure consisting of a vertical line of length $1-\beta \in (0,1)$
symmetrically placed in a rectangle with width $L \in (0,\infty)$ and
normalized height 1, see \fig~\ref{fig:system}(a). 
The special case $\beta = 1/2$ corresponds to the usual barrier
billiard~\cite{HM90}. 
The trivial energy dependence of the system is ignored by setting the
energy to 1/2, or equally the magnitude of the momentum $(p_x,p_y)$ to
1. Given an initial momentum only a finite number of directions is realized
during the time evolution, due to the fact that all angles in the polygon are
rational multiples of $\pi$. In other words, the motion in phase space takes
place on two-dimensional invariant surfaces. Using the general formula for the
genus of such surfaces~\cite{RichensBerry81} gives~2, i.e. the surfaces have
the topology of two-handled spheres and not that of tori (single-handled
spheres). 
\def\figsystem{%
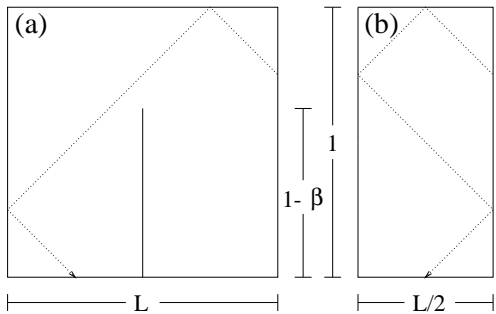
(a) Barrier billiard, rectangle with a vertical
line connecting the origin of the coordinate system $(x,y) = (0,0)$ with the
point $(x,y) = (0,1-\beta)$. (b) Symmetry reduced version.}
\def\FIGsystem{\centerline{\psfig{figure=system.eps,width=6.5cm,angle=0}
\vspace{0.25cm}
}}
\FIGo{fig:system}{\figsystem}{\FIGsystem}

Our model satisfies the definition of almost-integrability given by
 Gutkin~\cite{Gutkin86} for all values of $L$ and $\beta$. 
It is composed of two copies of an integrable rectangular billiard, which can
 be obtained by symmetry reduction, see \fig~\ref{fig:system}(b).
The dynamics can be split into two parts. 
First, the quasiperiodical dynamics $(|x(t)|, y(t))$ on the invariant tori of
the symmetry reduced system.
Second, the dynamics of the sign of $x$, $s(t) = \pm 1$. 

The quasiperiodical component is most elegantly described 
in action-angle variables. The actions $I_x = |p_x|L/2\pi$ and $I_y =
|p_y|/\pi$ label the invariant tori while the angles (divided by $2\pi$ and
computed modulo 1)
describe the flow on a given torus 
\bega
\label{eq:phixt}
\phi_x(t) & = & \frac{\omega_x}{2\pi}t+\phi_{x,o} \quad (\mbox{mod}\; 1) \\
\label{eq:phiyt}
\phi_y(t) & = & \frac{\omega_y}{2\pi}t+\phi_{y,o} \quad (\mbox{mod}\; 1),  
\enda
with the frequencies $\omega_x = 4\pi^2I_x/L^2$ and $\omega_y =
\pi^2I_y$. The flow on the torus is ergodic if and only if the winding number
$\rho = \omega_y/\omega_x = I_y L^2/(4I_x)$ is irrational. 
The motion in configuration space is explicitly given by
\bega\label{eq:xt}
|x(t)| & = & \frac{L}2f(\phi_x(t)) \\
\label{eq:yt}
y(t) & = & f(\phi_y(t)+\frac{1-\beta}2 \quad (\mbox{mod}\; 1)) 
\enda
with the piece-wise linear function 
\[
f(u) = \Biggl\{\begin{array}{cl}  
2u & \mbox{if} \quad 0 \leq u < 1/2 \\ 
2(1-u) & \mbox{otherwise.} \end{array}
\]
The shift in the argument of the function $f$ in \equ~(\ref{eq:yt}) will be
convenient later.

Now we consider the non-quasi-periodical part of the dynamics.
Clearly, the value of $s(t)$ changes when the trajectory impinges on the
segment \{$x=0$ and $1-\beta < y$\} (or \{$\phi_x = 0$ and $0 < \phi_y <
\beta$\}). We therefore introduce the Poincar\'e section $\phi_x(nT) = 0$ with
$T = 2\pi/\omega_x$ and the integer $n$. We obtain 
\bega\label{eq:map1}
\phi_{y,n+1} & = & \phi_{y,n} + \rho  \quad (\mbox{mod}\; 1) \\
\label{eq:map2}
s_{n+1} & = & s_n \Phi(\phi_{y,n})
\enda
with
\[
\Phi(\phi) = \Biggl\{\begin{array}{cl}  
-1 & \mbox{if} \quad 0 < \phi < \beta \\ 
1 & \mbox{otherwise.} \end{array}
\]
Note that diffractive orbits, i.e. orbits which hit the edge of
the barrier, can be neglected since they have measure zero in phase space. 
The two-dimensional mapping~(\ref{eq:map1})-(\ref{eq:map2}) has the solution
\bege\label{eq:formalsolution}
s_n = s_0 \Pi^n_{j=0} \Phi(\phi_{y,0}+j\rho \; (\mbox{mod}\; 1)) .
\ende
This equation together with \equ~(\ref{eq:phixt})-(\ref{eq:yt}) and $n$ given
by the integer part of $t/T$ solve the billiard problem. But the
solution is only formal, because in order to compute~(\ref{eq:formalsolution})
one has to iterate~(\ref{eq:map1})-(\ref{eq:map2}).
A similar, but more complicated, solution of the rhombus billiard was reported
in~\cite{EFV84}. But its formality was not recognized since the authors
focussed only on the equation corresponding to~(\ref{eq:map1}), while the
equation corresponding to~(\ref{eq:map2}) was hidden in the notation.  


Having defined the dynamical system, we now discuss the spectral properties
of a typical smooth phase space function $F(x,p_x,y,p_y)$ (nontypical
functions $F(x^2,p_x^2,y,p_y)$ obviously have discrete spectra). We
investigate firstly the function $F(t) = x(t)$ (the argumentation for $p_x(t)$
is identical). We will see that $x(t)$ has a singular continuous spectrum,
which implies that a typical function has a mixture of discrete and singular
continuous spectrum. With $x(t) = s(t)|x(t)|$ it is clear that the Fourier transform of $x(t)$, 
\bege\label{eq:X}
X(\omega) = \int_{-\infty}^\infty x(t) e^{-i\omega t} dt \ ,
\ende
can be written as convolution of the Fourier transforms of $|x(t)|$ and $s(t)$,
$\bar{X}(\omega)$ and $S(\omega)$ 
\bege\label{eq:convolution}
X(\omega) = \int_{-\infty}^\infty \bar{X}(\omega')S(\omega-\omega')d\omega'.
\ende
Since $|x(t)|$ is periodic with period $T$ we have
\[
\bar{X}(\omega') = \sum_{m=-\infty}^\infty  A_m\delta(\omega'-m\omega_x) .
\]
Substituting into \equ~(\ref{eq:convolution}) gives
\bege\label{eq:Xresult}
X(\omega) = \sum_{m=-\infty}^\infty  A_mS(\omega-m\omega_x) .
\ende

Since $s(t)$ changes in discrete steps it is suitable to
consider the discrete-time Fourier transform of the sequence~$[s_n]$ 
\[
\Sigma(\omega) = \sum_{n=-\infty}^\infty s_n e^{-i\omega n} 
\]
which is related to the continuous-time Fourier transform via
\bege\label{eq:cdFt}
S(\omega) = \frac{1}{i\omega}(1-e^{i\omega T})\Sigma(\omega T) .
\ende
Note that the prefactor goes to the constant $T$ as $\omega \to 0$.
We can now take advantage of the fact that the
mapping~(\ref{eq:map1})-(\ref{eq:map2}) is well studied. It is proven (see
\cite{Riley78} and references therein) that if $\rho$ is irrational then
$\Sigma(\omega)$ is singular continuous for almost all $\beta$ (the spectrum is
discrete, e.g., if $\beta = n\rho\; (\mbox{mod}\; 1)$).   
Since this holds trivially also for any nonconstant function of $s(t)$ we
conclude that the $s$-motion is weakly mixing~\cite{CFS82}.
The fact that almost all trajectories in our billiard have irrational winding
number $\rho$, independent of the values of the parameters $L$ and $\beta$,
ensures that typical trajectories give rise to singular spectra
$\Sigma(\omega)$. Together with \equ~(\ref{eq:Xresult}) and (\ref{eq:cdFt})
this implies that $X(\omega)$ is typically singular continuous. Nevertheless,
there exist nonconstant functions of $x(t)$, e.g. $x^2(t)$, which do 
not have a singular continuous spectrum, so the $x$-motion cannot be weakly
mixing. This result indicates that a purely numerical detection of
weak mixing by investigating the spectral properties of just a
single phase space variable may lead to wrong conclusions. 

Figure~\ref{fig:fourier} shows an example of $|X(\omega)|^2$ close to the
frequency $\omega_x$ (so essentially $|S(\omega-\omega_x)|^2$ is
shown) for the case $\beta = 1/2$ and $\rho$ equal
the reciprocal of the golden mean, $\rhogm = (\sqrt{5}-1)/2$. The
integration~(\ref{eq:X}) has been performed with a fading factor
$\exp{(-2t/t_{\text {max}})}$ and truncated at $t_{\text {max}} = 10000$.  
The enlargement in \fig~\ref{fig:fourier} reveals some selfsimilarity of
$S(\omega)$ for this set of parameters. The minor lack of symmetry is due to the
prefactor in \equ~(\ref{eq:cdFt}); $\Sigma(\omega)$ is
perfectly selfsimilar as shown by Feudel et al.~\cite{FPP96} by using a
renormalization group approach (compare the magnification in
\fig~\ref{fig:fourier} with \fig~3 in \cite{FPP96}). 
Feudel et al. have also calculated the generalized dimensions: 
(i) the Hausdorff dimension $D_0 = 1$, i.e. the spectrum is continuous.
(ii) The information dimension $D_1 \approx 0.75$, i.e. the spectrum
is singular. 
(iii) The correlation dimension $D_2 \approx0.65$, which is the exponent of
the power-law decay $t^{-D_2}$ of the integrated AF~\cite{KPG92}.
\def\figfourier{%
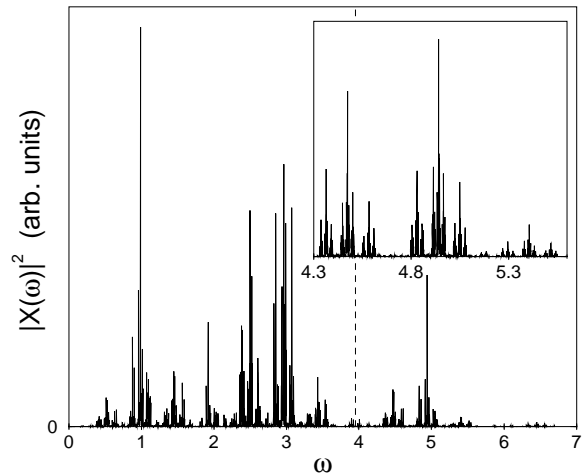
Power spectrum $|X(\omega)|^2$ for $\beta = 1/2$ and $\rho = \rhogm$.
The dashed line marks the fundamental frequency of $|x(t)|$, $\omega_x \approx
3.95187$. The inset shows a magnification.} 
\def\FIGfourier{\centerline{\psfig{figure=fourier.eps,width=7.5cm,angle=0}
}}
\FIGo{fig:fourier}{\figfourier}{\FIGfourier}

Before discussing the integrated AF, let us first consider the (normalized) AF
($x(t)$ has zero mean value)
\bege\label{eq:af}
R_x(\tau) = \frac{\langle x(t+\tau)x(t)\rangle}{\langle x^2(t) \rangle} 
\ende
where the average $\langle \ldots \rangle$ is taken over $t$.
We restrict ourself to the times $\tau = nT$. By using the periodicity of
$|x(t)|$ it is an easy matter to show that $R_x(nT)$ is equal to $R_s(n)$, the
AF of the process $s_n$ (where the average is over the discrete time
$n$). This means we discuss here the correlations of the transitions between
the two copies of the integrable subbilliard.  
For the special case $\beta = 1/2$, $R_s(n)$ as function of $n$ is studied
in~\cite{PikovskyFeudel94b} for different $\rho$.
It is shown that $R_s(n)$ vanishes for odd $n$. For large even $n$ it is mostly
close to zero but has large peaks at positions which roughly grow
exponentially. For example, if $\rho = \rhogm$ then peaks with amplitude
$\approx 0.55$ are located at all even Fibonacci numbers surrounded by smaller
peaks in a selfsimilar manner. 

We have determined the integrated AF 
\begin{equation}\label{eq:intAF}
R_{s,{\text {int}}}(n) = \frac1n\sum_{j=0}^n |R_s(j)|^2 
\end{equation}
up to $n_{\text {max}} = 10000$. Its asymptotic behaviour is found to be in
 agreement with the expected power-law decay, which allows us to compute $D_2$
 numerically.
For fixed $\rho = \rhogm$, $D_2$
as function of $\beta$ (using 1000 randomly distributed $\beta$-values) has
 mean value $0.458$ and shows very strong fluctuations, ranging from large
 decay rates $D_2 = 0.68$ to zero decay. These fluctuations are by no means
 surprising, since there are dense sets of $\beta$-values (e.g. $\beta =
 n\rho\; (\mbox{mod}\; 1)$) for which $D_2$ vanishes.   
We have also calculated $D_2$ for fixed $\beta = 1/2$ as a function of
$\rho$ (using 1000 $\rho$-values). Here, $D_2$ is more or less uniformly
distributed between zero and $0.94$ around a mean value of $0.469$.
Due to the sensitive dependence of $D_2$ on the winding number, it seems
useful to perform an average of the AF over all invariant surfaces (i.e. over
all $\rho$) in order to characterize the correlation properties of the
transitions between the two copies of the subbilliard as a whole. But this
kind of averaging does not yield a reasonable correlation function since for
an integrable system such a function would indicate correlation
decay~\cite{CV81}.   

The fact that the integrated AF~(\ref{eq:intAF}) decays to zero but $R_s$
does not confirms the weak-mixing property of the $s$-motion. Because of
$R_x(nT) = R_s(n)$ one could naively draw the wrong conclusion that the
$x$-motion is also weakly mixing. This confusion can be resolved if one clearly
distinguishes between continuous-time and discrete-time approaches. 
Trivially, $R_x$ does not decay for large continuous times $\tau$ if it
does not decay for large discrete times $n$. But we expect that the integrated
(now really an integration not a summation) AF of $x(t)$ does not decay,
 in contrast to the integrated AF of the process $s_n$ (the integrated AF of a general function $F(x,y,p_x,p_y)$ obviously shows zero
decay rates, since $y(t)$ is periodic). This agrees with results of the study
on an almost-integrable version of the ``square-ring billiard''~\cite{AGR99}. 
It is worth emphasising that the discrete-time approach, which is always used
in numerical studies where integrations are replaced by summations (for
billiards the discrete-time approach is often used explicitly), can
wrongly indicate weak mixing, even though this is quite unlikely.  

In terms of the billiard motion we have the following scenario. The motion is
not mixing and even not weakly mixing; that means, a small cluster of
particles with the same initial momenta will not spread out on the entire
billiard table.   
But the motion is ``partly weak mixing'' in the sense that after some transient
time (when the AF is close to zero) there are two clusters, one in the
left and one in the right part, each of them consisting of approximately half
of the particles.   
At ``resonant'' times (corresponding to the peaks of the AF) one of the two
clusters contains more particles than the other one. But these reclusterings
become more and more rare with increasing time.

\rem{It is worth noting that the system can easily generalized to an almost
integrable system whose invariant surfaces have arbitrary genus by ``cutting
holes'' in the barrier. The overall description of the dynamics remains the
same, only the function $\Phi(\phi)$ becomes more complicated. The qualitative
correlation properties stay also the same (one small cluster spread into two
clusters). Numerical simulations indicate that the quantitative correlation
properties depend strongly on the position of the holes rather than of the
number of it. 
A far more dramatic change in the dynamics can be achieved by shifting the
barrier horizontally rather than by cutting holes in it. If we put the barrier
on a rational $x/L \neq 0$ than a small cluster can spread into more than two
clusters. 
We believe that if we lift the almost-integrability by placing the barrier on
an irrational $x/L$ than the $x$-motion would be weakly mixing (the $y$-motion
is still periodic).  
This contradicts the naive point of view that the higher the genus of the
invariant surfaces the more irregular the motion.   
}
To summarize, we have proven that the motion in the pseudo-integrable barrier
billiard exhibits a discrete and a singular continuous spectral component. We
expect that this is true for all almost-integrable billiards.

I would like to thank R. Artuso for useful comments and 
M. Sieber and J. N{\"o}ckel for discussions. 
	
\bibliographystyle{prsty}
\bibliography{}
\clearpage
\end{multicols}
\end{document}